\documentclass[11pt,oneside,letterpaper]{article}
\usepackage{graphicx,amssymb,amsmath,anysize,epsfig}

\begin{document}
\title{\bf Comment on ``Preacceleration without radiation:\\
 The nonexistence of
preradiation phenomenon," \\by J. A. Heras [Am. J. Phys. 74, 1025
(2006)]}
\author{V. Hnizdo$^{\rm a)}$\\
\it{National Institute for Occupational Safety and Health,
Morgantown, West Virginia 26505} }
\date{}
\maketitle

In a recent paper,$^1$  Heras concludes that there is no radiation
in the preacceleration parts of non-runaway solutions of the
Abraham--Lorentz (AL) equation of motion of a point charge. The
purpose of this comment is to argue that this conclusion is
incorrect.

The non-runaway solutions of the AL equation  have a well-known
unphysical feature of preacceleration: the charge starts to
accelerate already at a time of order $\tau= 2 e^2/(3mc^3)$, where
$e$ and $m$ are respectively the charge's magnitude and mass, {\it
before} an external force is applied. Heras studies the non-runaway
solutions of the AL equation for the class of external forces that
act only in a finite interval of time $(0,T)$, and reaches the
conclusion that, despite its nonzero preacceleration, the charge
does not radiate at times $t<0$. He bases this conclusion on an
integration-by-parts transformation of the integral expression for
the energy $W_R(0,T)$ radiated according to the Larmor radiation
rate during the time interval $(0,T)$ into a sum of two terms, one
of which  is an integral over the interval $(0,T)$ and the other
equals the negative of the energy $W_R(-\infty,0)$ that is radiated
according to the Larmor formula in the interval $(-\infty,0)$, i.e.,
during the preacceleration. This transformation leads to the
mathematical cancelation of the energy $W_R(-\infty,0)$ in the total
radiated energy $W_R(-\infty,0)+W_R(0,T)$ by the equal and opposite
term $-W_R(-\infty,0)$ in the energy $W_R(0,T)$,  and thus the total
radiated energy can be expressed as an integral over only the time
interval $(0,T)$.  Heras argues against ``a misleading time
separation of [the radiated] energy in the interval of
preacceleration $(-\infty,0)$ and the interval $(0,T)$" because the
total radiated energy can be expressed as an integral over only the
interval $(0,T)$, which ``shows that the energy is radiated only
during $(0,T)$, which is the interval when the force acts."

To strengthen this argument,  using the AL equation Heras writes the
Larmor radiation rate $m\tau \dot{v}^2$ as
$m\tau\dot{v}^2-m\tau^2\dot{\bf v}{\bf\cdot}\ddot{\bf
v}+m\tau^2\dot{\bf v}{\bf\cdot}\ddot{\bf v} =\tau \dot{\bf v}{\bf
\cdot F}+ (d/dt)(\frac{1}{2} m\tau^2\dot{v}^2)$, where $\bf F$ is
the external force, which gives the total radiated energy
$W_R(-\infty,\infty)$ as $\tau \int_0^T \dot{\bf v}{\bf \cdot
F}\,dt$ because the force is assumed to act only in the interval
$(0,T)$ and the integration over the time-derivative term must
vanish for non-runaway solutions that are subject to the boundary
condition $\dot{\bf v}(\pm\infty)=0$. This expresses the total
radiated energy again as an integral over only the interval in which
the external force acts.

However, expressing the total radiated energy as an integral over
the time interval in which the external force is nonzero does not
necessarily mean that this energy is  radiated only in that
interval. The total energy radiated over a nonvanishing interval of
time can be expressed as a sum of various terms in many different
ways that do not necessarily associate the correct time origins to
the terms, without affecting the {\it global} energy balance over
that time interval. But the AL equation implies also a strict {\it
instantaneous} power balance$^2$
\begin{equation}
{\bf F \cdot v} = \frac{d}{dt}\,({\textstyle \frac{1}{2}}m v^2) +m
\tau \dot{v}^2 -\frac{d}{dt}\, (m\tau {\bf v\cdot} \dot{\bf v}).
\end{equation}
Here, the left-hand side is the rate of  work of the external force
$\bf F$ on a charge moving with velocity $\bf v$, and the three
terms on the right-hand side are respectively the rate of change of
the charge's kinetic energy, its Larmor radiation rate, and the
negative of the rate of change of the so-called Schott energy, which
is a reversible energy in the charge's near electromagnetic field.
The balance (1) must be satisfied at any instant $t$, including
those of any preacceleration time interval, when ${\bf F}=0$  and
thus ${\bf F \cdot v}=0$, but all the three terms on the right hand
side of (1) are nonzero so that their sum vanishes. This is easily
verified with the $t<0$, i.e., preacceleration, values of $v(t)$ and
$a(t)=\dot{v}(t)$ in the example solutions (13), (16), (20) and (23)
of Heras.  The Schott-energy term  is crucial for maintaining  the
instantaneous energy conservation during preacceleration; Heras
himself has co-authored a recent paper that emphasizes the role of
this term in the instantaneous power balance of a nonrelativistic
classical charge.$^2$

Using the instantaneous power balance (1), one can express the total
radiated energy for the non-runaway solutions  considered by Heras
as
\begin{equation}
 W_R(-\infty,\infty)\equiv \int_{-\infty}^{\infty} m\tau \dot{v}^2\,dt
=\int_0^T{\bf F \cdot v}\,dt- ({\textstyle
\frac{1}{2}}mv_{\infty}^2-{\textstyle \frac{1}{2}}mv_{-\infty}^2),
\end{equation}
as the external force $\bf F$ is nonzero only in the time interval
$(0,T)$ and the term $m\tau {\bf v\cdot} \dot{\bf
v}|_{-\infty}^{\infty}$ vanishes because of the boundary condition
$\dot{\bf v}(\pm\infty)=0$.  Equation (2) expresses the fact that
the total work $\int_0^T{\bf F \cdot v}\,dt$ of the external force
equals the total radiated energy plus the total increment
${\textstyle \frac{1}{2}}m v^2_{\infty}-{\textstyle \frac{1}{2}}m
v_{-\infty}^2$ in the charge's kinetic energy, but gives no
indication of the time interval in which any part of the energy
$W_R(-\infty,\infty)$ is radiated.

All solutions of the AL equation  satisfy the instantaneous power
balance (1) because this balance is demanded by the AL equation
itself. This includes the grossly unphysical runaway solutions,
which on a first sight would seem to violate energy conservation,
and the uniformly accelerated motion driven by a constant external
force, where the AL radiation-reaction force vanishes despite the
fact that the charge radiates according to the Larmor rate.$^3$
Solutions that feature preaceleration are no exception. The
instantaneous power balance implied by the AL equation unambiguously
demands that even a ``preaccelerating" charge radiates according to
the Larmor radiation-rate formula.

Heras's conclusion that there is no radiation in the preacceleration
part of a non-runaway AL trajectory amounts to a claim that the
Larmor radiation rate is not applicable on that part of the
trajectory. But there are no grounds within nonrelativistic
classical electrodynamics for denying the applicability of the
Larmor formula to the calculation of  the energy radiated by an
accelerating charge even over a time interval as short as the time
$\tau= 2 e^2/(3mc^3)$
--- if the idealization of a strictly point charge is assumed. The
AL equation assumes this idealization, and all indications are that
already the very concept of a point charge (or, more generally,  a
charge the spatial extension of which is smaller than the classical
radius $r_c=e^2/mc^2$ corresponding to its magnitude $e$ and mass
$m$) is in classical electrodynamics {\it unphysical} because
alternatives to the AL equation that explicitly or effectively
assume a charge of spatial extension of at least the order of $r_c$,
like the classical Ford--O'Connell equation,$^4$ or the
Landau--Lifshitz (LL) equation,$^5$ advocated by Rohrlich$^6$ as the
correct equation of motion of a classical point-{\it like}
charge,$^7$ do not admit solutions with unphysical features as
preacceleration or runaway behavior. The radiation rate implied by
the nonrelativistic limit of the LL equation is $\tau \dot{\bf v}\bf
{\cdot F}$,$^8$ which shows clearly that the charge then radiates
only when the external force $\bf F$ does not vanish.
\\

\noindent $^{\rm a)}$Electronic mail: vbh5@cdc.gov. V. Hnizdo has
written this comment in his private capacity. No official support or
endorsement by Centers for Disease Control and Prevention is
intended or should be inferred.
\\
\noindent$^1$J. A. Heras, ``Preacceleration without radiation: The
nonexistence of preradiation phenomenon,"  Am. J. Phys. {\bf 74},
1025--1030 (2006).
\\
\noindent$^2$J. A. Heras and R. F. O'Connell, ``Generalization of
the Schott energy in electrodynamic radiation theory," Am.\ J.\
Phys. {\bf 74}, 150--153 (2006).
\\
\noindent$^3$The ``paradox" that a uniformly accelerated charge (or,
relativistically, a charge in ``hyperbolic" motion) radiates while
its self-force (the ``radiation-reaction" force)  vanishes has a
long history, but it is now almost universally accepted that a
uniformly accelerated charge radiates, see the recent paper  C. de
Almeida and A. Saa, ``The radiation of a uniformly accelerated
charge is beyond horizon: A simple derivation," Am.\ J.\ Phys.\ {\bf
74}, 154--158 (2006) and references therein. Curiously, in their
recent paper on the Schott energy (Ref.\ 2), Heras and O'Connell
assert that there is no radiated energy when the radiation-reaction
force vanishes, despite the fact that the presence of the Schott
term in the instantaneous power balance resolves any possible
difficulty with maintaining instantaneous energy conservation; a
constant acceleration, $\dot{\bf v}= {\rm const}$, only leads to the
last two terms in Eq.\ (1) canceling each other while the radiation
rate $m\tau\dot{v}^2$ does not vanish.
\\
\noindent$^4$G. W. Ford and R. F. O'Connell, ``Radiation reaction in
electrodynamics and the elimination of runaway solutions," Phys.\
Lett.\ A {\bf 157}, 217--220 (1991).  The nonrelativistic limit of
the Landau--Lifshitz equation (see Refs.\ 5 and 6) is a special case
$\Omega=1/\tau$ of the classical Ford--O'Connell equation, where
$\Omega$ is a frequency cut-off parameter.
\\
\noindent$^5$L. D. Landau and E. M. Lifshitz, {\it The Classical
Theory of Fields} (Pergamon, Oxford, 1975), Sec.~76. Generally, the
LL radiation-reaction 4-force is that of the Lorentz--Abraham--Dirac
equation for a point charge in which the 2nd-order derivative
$\ddot{v}^{\mu}$ of the 4-velocity $v^{\mu}$ is replaced by
$\dot{F}^{\mu}/m$, where $F^{\mu}$ is the external 4-force, and
$\dot{v}^{\mu}$ in $\dot{F}^{\mu}$ is replaced by $F^{\mu}/m$ when
$F^{\mu}$ depends on $v^{\mu}$.
%
\\
\noindent$^6$F. Rohrlich, ``The correct equation of motion of a
classical point charge," Phys.\ Lett.\ A {\bf 283}, 276--278 (2001);
``Dynamics of a classical quasi-point charge," Phys.\ Lett.\ A {\bf
303}, 307--310 (2002).
\\
\noindent$^7$The LL radiation-reaction 4-force has been derived
recently assuming a quasi-rigid particle of radius $r\ge
2e^2/(3mc^2)$ and the condition $|\dot{F}|/F\ll 1/\tau$ on the
external force, see R. Medina, ``Radiation reaction of a classical
quasi-rigid extended particle," J.\ Phys.\ A {\bf 39}, 3801--3816
(2006).
\\
\noindent$^8$The radiation rate $\tau \dot{\bf v}\bf {\cdot F}$ and
a Schott energy $\tau \bf{v\cdot F}$ are the correct nonrelativistic
limits of the corresponding quantities implied by the LL equation;
{\it cf}.\ the values $P_{\rm FO}=(\tau/m)F^2$ and $E_{\rm
FO}=\tau{\bf v\cdot F}-(\tau^2/2m)F^2$ used with the classical
Ford--O'Connell equation in Ref.\ 2.

\end{document}